# *Panel*: Economic Policy and Governance during Pandemics using AI

**Feras A. Batarseh**
Commonwealth Cyber Initiative
Virginia Tech
Arlington, VA 22203
batarseh@vt.edu

**Munisamy Gopinath**
Department of Agricultural and Applied Economics
University of Georgia
Athens, Georgia 30602
m.gopinath@uga.edu

Abstract – The global food supply chain –starting at farms and ending with consumers– has been seriously disrupted by many outlier events such as trade wars, the China demand shock, natural disasters, and pandemics. Outlier events create uncertainty along the entire supply chain in addition to intervening policy responses to mitigate their adverse effects. Artificial Intelligence (AI) methods (i.e. machine/reinforcement/deep learning) provide an opportunity to better understand outcomes during outlier events by identifying regular, irregular and contextual components. Employing AI can provide guidance to decision making suppliers, farmers, processors, wholesalers, and retailers along the supply chain, and policy makers to facilitate welfare-improving outcomes. This panel discusses these issues.

## Motivation and Background

In recent years, the global supply chain has witnessed significant disruptions. While natural disasters commonly affect this chain, climate change may be adding to their rigor and frequency. Moreover, a trade war between the U.S. and China and the recent Covid-19 pandemic have created to additional uncertainties in global production chains and irregularities in the supply and demand of essential goods.

The three most critical adverse world incidents since the 1870s were World War II, the Great Depression in America, and World War I. Results from multiple studies suggest that the Great Influenza Pandemic of 1918-1920 is the next most important negative economic shock for the world (NBER, 2020). Modeling and predicting the implications of such outlier events is an important endeavor. While traditional economic models aim to be reliable predictors, we consider the possibility that AI methods allow for better predictions, outlier detections, and classifications to inform policy and decisions along the supply chain. For instance, data analytics and cyber tools for real time decision making can help farmers in deploying quick decisions for improved precision agriculture.

During these outlier events, analysis is performed in uncharted waters. Issues arise such as the need to use daily if not hourly data for pattern recognition and predictions. Additionally, decisions need to be executed in a quick manner using real time analysis and on-demand analytics.
We aim to discuss data platforms and tools that enable the execution and development of AI methods such as: Deep Learning (DL) through neural networks for predictions and outlier detection, Reinforcement Learning (RL) for causality and patterns recognition, as well as Machine Learning (ML) for forecasts and associations. The goal is to provide an open-source data system that is hosted through governmental FAIR (Findable, Accessible, Interoperable, and Reusable) means. This panel is relevant to three categories of end users: 1. AI practitioners, 2. policy makers, and 3. Decision-makers along the supply chain.

## Pandemics and Ag Economics

In our work, we initially focus on pandemics as an example outlier event; and we model effects on the food supply chain accordingly. Models imply that people's decision to cut back on consumption away from home reduces the severity of the pandemic, as measured by total infections and deaths (Eichenbaum et al., 2020). These decisions however, exacerbate the size of consumption at home and the consequent economic impacts of the pandemic. Tradeoff analysis is critical to balancing pandemic response and economic outcomes. If the infected people do not fully internalize the effect of their economic decisions on the spread of the virus, the competitive equilibrium may not be socially optimal. So how should farmers deal with such drastic shifts in demand? and how should retailers handle the several-fold surge in demand? Getting the right information to end users and decision makers is critical (to avoid *infodemics*), that is best done through agriculture cyber-informatics tools that are driven by AI.

Real time policy- and decision-making during outlier events such as pandemics are challenging tasks. For instance, during pandemics, premature deaths reduce the labor force; illness leads to absenteeism from work and eventually to reduced overall productivity of a nation. Economic resources flow to treatment and control measures to reduce disease spread to avoid a serious disruption to economic activity (Fan et al., 2020).

During the Covid-19 outbreak, most models from the Centers for Disease Control and Prevention (CDC) and the World Health Organization (WHO) were inconsistent with reality (predictions were very different than actual values) (CDC, 2019). The reason being they were modeled without looking into the predictions through the lens of multivariate *outliers* (Batarseh et al., 2019). Albeit most traditional economic models deploy causality analysis to infer policy, regular models do not account much for outliers (Batarseh and Yang, 2020), and most analysts deal with outliers based on human experience or elimination of such events in models (Batarseh et al., 2017).

## AI and Covid-19 Policy Implications

In this panel, we propose methods to evaluate shocks (such as pandemics and the China shock) and support real time analysis for decision makers. An important dimension of the cost of a pandemic lies in its impact on income, consumer behavior, unemployment, mortality rates, production, and multiple other effects. Such factors ought to be modeled so better real time decisions are presented with more evidence and data-driven reasoning. But not all outliers are created equal; the historical record suggests that the 1918 influenza was an outlier among outliers, with unusual circumstances including the co-occurrence of World War I. No other influenza pandemic on record had such devastatingly high mortality rates, with estimates ranging from 20 to 50 million excess deaths over the period 1918-20 (Fan et al., 2020). Prior to Covid-19, researchers at the CDC calculate (using traditional models) that deaths in the United States could reach 207,000, and the initial cost to the economy could approach $166 billion, or roughly 1.5 percent of GDP in case of an international pandemic similar to the 1918 outbreak (Garrett, 2008).

The Covid-19 crisis erupted and unfolded with tremendous speed. Take the U.S. case as an example: in February 2020, the unemployment rate was 3.5%, equaling its lowest rate in the past ~67 years. Only six weeks later, the outlook has shifted profoundly: nearly eleven million Americans filed for unemployment benefits in two weeks (Chaney and Morath, 2020). Millions more lost jobs but did not file. Because the outlook changed with such suddenness, methods based on backward-looking statistical analyses and historic data are unlikely to yield suitable measures of forward-looking uncertainty.

Timeliness of data is a critical practical challenge as well. To estimate the current and future macroeconomic effects of Covid-19 induced uncertainties, we need measures that are available in real time, or nearly so. While all industries have been seriously affected by the pandemic, food and agriculture have been among the hardest hit segments of the U.S. economy. The primary reason lies in the composition of household food expenditures. Nearly 55% of food consumed away from home (restaurants, ball parks, bars and pubs) had to be routed through grocery stores and food delivery services. The impacts of the pandemic appear to vary by commodity based on two critical issues: perishability and labor use. Perishables like fruits, vegetables and milk are among the hardest hit. Many of these industries also depend on labor for growing and harvesting. While there is no immediate shortage of food in the U.S., according to the U.S. Department of Agriculture (USDA), however the current demand for items such as grocery-size products and on-demand delivery is greater than what is in abundant supply: bulk, large-sized products and processed shipments to restaurants that remain open. This demand-supply mismatch appears to mimic anecdotal evidence of price spikes and empty store shelves on the consumer side and the collapse of demand and dumping of food on the farm side, with a range of linked effects in the middle.

The results include a significant slowdown in the operations of processing and distribution, shortage of workers at farm, processing and distribution (trucking) facilities, and a shortage of cleaning and sanitizing supplies. Compounding these effects as the resources spent in contact tracing and quarantine if/when a worker tested positive for Covid-19.

Additionally, there is an urgent need to transition products that run through food services into those that consumers need at the grocery store. For example, restaurants usually bought diced vegetables, like onions, in 60-pound bags, but consumers at grocery stores usually buy 3- to 5-pound bags of unpeeled onions. Also, large cheese blocks sold to food services, which generally have sizable storage space, cannot be chopped overnight into packs of ounces and pounds to sell at grocery stores. AI tools are critical in measuring the effects of this pandemic on the supply chain to guide real-time decision-making. Moreover, if we know how outlier events affect the economy, policy makers can redefine their models to mitigate the adverse effects.

## Datasets and Methods

Our work leverages agricultural commodity data at the annual, quarterly, monthly, weekly and daily levels, from USDA, the World Bank (WB), International Monetary Fund (IMF); and other sources such as the World Trade Organization's (WTO) GATT system for trade and commerce data. Additionally, we deploy AI paradigms including ML models and their refinements (such as: classification and boosting), DL methods such as neural networks, RL methods such as Q-learning, and other potential AI methods such as contextual AI, and causal learning.

For objectives of this work, comparisons of AI-based forecasts with currently available approaches – linear, factor and non-linear models and Bayesian approaches – will be made. Futures markets are an example existing benchmark system for forecasting prices, however, they only exist for certain commodities, locations, and time periods. Additionally, many commodities have no futures markets; for example, in the U.S., sorghum, grain, barley, canola, and other commodities. We aim to address this gap. ML, RL and DL algorithms permit model and data "drifts," essentially presenting opportunities to readjust the models and produce alternative results for evaluation.

ML methods will be used (based on the datasets collected from the sources mentioned above) for predictions of economic variables such as prices, tariffs, production, and utilization. For instance, RL is tested for finding the optimal scale of consumption and production of a commodity during conventional times vs. during shocks. Using reward and punishment of the RL algorithm, the reward would recommend methods for lower prices for consumers and/or higher profits for producers while considering tens of other economic variables. Such optimizations and equilibriums aid in policy scenario management and analysis at the government (federal and state levels).

## Discussions and Conclusions

AI methods can point to such issues and provide solutions to the mitigation of imbalance, bias, and outliers – especially for economic applications. The 2020 U.S. National Artificial Intelligence Initiative Bill states: "Artificial intelligence is a tool that has the potential to change and possibly transform every sector of the United States economy" (S.1558 – AIIA, 2020). Few studies have applied AI to economics; accordingly, some have called for increased attention to the role of AI tools (Mullainathan and Spiess, 2017). Also, a study by NBER points to the potential of economists adopting AI and ML more than any other empirical strategy (Athey, 2019). Open-government data provide the fuel to power the mentioned methods. Unlike traditional approaches, e.g. expert-judgment based linear models or time-series methods, AI methods provide a range of data-driven and interpretable projections for farm and policy decisions. Prior studies indicate the high relevance of AI methods for predicting a range of economic relationships with a greater accuracy than traditional approaches (Athey and Imbens, 2017). A very distinct difference between AI methods and econometrics is that the latter aims to identify causalities, while the former is concerned with regressions, classifiers, clusters, associations, optimizations and multiple other actionable outcomes.

In 2013, Gevel et al. published a book called "the Nexus between Artificial Intelligence and Economics". It was one of the first few works that introduced agent-based computational economics (Gevel et al. 2013). One year later, Feng et al. (2014) studied economic growth in the Chinese province of Zhejiang using a neural networks model. Their method however, is very limited in scope, and proves difficult to deploy across other provinces in China or other geographical entities in other countries. Abadie et al. (2010) developed a similar model, but applied it to the rising tobacco prices in California. In 2016, Milacic et al. (2016) expanded the scope, and developed a model for growth in GDP including its components: agriculture, manufacturing, industry, and services. Falat et al. (2015) developed a set of ML models for describing economic patterns, but did not offer predictions.

None of the mentioned studies evaluates outlier events such as pandemics on the economy, or how that affects supply chains and overall policy.

The results of AI endeavors are majorly driven by data quality. Throughout these deployments, serious showstopper problems are usually unresolved, such as: data collection ambiguities, data imbalance, hidden biases in data, the lack of domain information, and data incompleteness (all under the umbrella of AI assurance). We are discussing an AI framework that considers such challenges and provides real time data analysis, especially during economic shocks.

## Panel Composition and Goals

This panel aims to address the following questions:

1- A discussion on Covid-19 and its economic modeling.
2- How can AI methods aid economic analysis in general?
3- How to use data and AI to model outlier events and their effect on the overall supply chain?
4- How are conventional statistical methods challenged during outlier events?
5- Context-based analysis of shocks and machine learning methods for management of unique contexts.
6- Use cases from the Federal Reserve Board, and U.S. Department of Agriculture.
7- How can we ensure real-time data collection to power AI-based analysis of the economic impact of pandemic, and to define policy implications?
8- How can AI methods be validated and verified for better policy making?

**Panel members:**
Panelists from Academia, Government, and Industry will be discussing this matter.

**Moderator:**
Feras A. Batarseh, Commonwealth Cyber Initiative, Virginia Tech, Arlington, VA.

**Panelists:**
Munisamy Gopinath, University of Georgia, Athens, GA.
Laura Freeman, Commonwealth Cyber Initiative, Virginia Tech, Arlington, VA.
Maros Ivanvic, Economic Research Service, U.S. Department of Agriculture, Washington, DC.
Anderson Monken, Federal Reserve Board and Georgetown University, Washington, DC.
Badri Narayanan Gopalakrishnan, Infinite Sum Modelling LLC, Seattle WA.